# Design and Realization of a Reflectarray Compact Antenna Test Range

You Wu, Zhiping Li∗, Jianhua Wu, and Zhengpeng Wang

*Abstract*— This paper presents a reflectarray compact antenna test range (RACATR). The new range layout is designed to suppress the unwanted diffraction. A fast near field optimization method has been used to determine the phase distribution of the only phased aperture efficiently in numerical example. The full wave simulation results of the linear reflectarray support that this facility design approach is working well to suppress the structural interference such as the mirror reflection and the edge diffraction. A 460 mm × 460 mm square reflectarray was manufactured and measured. Satisfying the requirement of the quiet zone field distribution, the quasi-plane wave is achieved from 26.5~28.5 GHz by linearly adjusting the feed location.

## I. INTRODUCTION

WHETHER for antenna measurements or radar cross section (RCS) measurements, the reflector Compact Antenna Test Range (CATR) is a mature and classical technology to provide a high accuracy plane wave in the near field [1]. The tight surface accuracy requirements limit the reflector application at high frequency, above 100 GHz, for example. Scanning synthetic aperture, lens, hologram, phased array and reflectarray antenna are the potential substitutes of the reflectors. Hologram CATR was developed to produce a plane wave for the millimeter and sub-millimeter wave antenna measurement successfully [2-3]. The space and energy efficiency of the hologram CATR is not high because the transmission type layout and the insertion loss of the hologram aperture [2-3]. Compared with other alternatives, the RACATR has many structural advantages of being planar, low-profile, lightweight, easy-to-fabricate, low cost and compact. However, the most severe disadvantage of the RA is the inherent feature of narrow bandwidth [4]. The wide band performance of RA is primarily limited by two factors. One is the narrow bandwidth of the patch elements and the other is the spatial phase delay varies with frequency. Fortunately, many wideband elements have been proposed to relieve the first factor, such as triple-resonant dipoles [5], double cross loops [6], windmill-shaped elements [7] and the Phoenix cell [8]. Moreover, a wideband method through linearly moving the feed location was described in [9] to compensate the different spatial phase delay for the hologram CATR. This method is based on the principle of electromagnetic similarity and could be applied for the improvement on the bandwidth of RA.

The reflectarray antenna is conventionally utilized in far field application such as beam shaping [10], multiple beam pattern [11], and contour beam applications [12]. In recent years, the near field applications of RA have been proposed for Radio-Frequency Identification (RFID) reader application [13]. However, generating the plane wave in the near field with the RA gets less attention. The concept that applying the reflectarray antenna in CATR system is proposed first in [14]. The simulation arrangement of RACATR in [14-18] is similar to the facility layout of the reflector CATR. However, the direct application of the facility layout of the reflector CATR in RACATR system brings a problem that the structural reflection will disturb the quiet zone field. The structural reflection can be attributed to the structural mode from the antenna scattering [19]. The new facility layout is proposed for preventing the desired plane wave field from the reflection interference as the main contribution in this paper. The goal of the research is to develop a new design technique of RACATR. The design will be a low-cost candidate solution for the millimeter wave antenna measurement of the 5G and the automotive radar in the future.

This paper is organized as follows. Section II describes the design approach and related theories. Simplified numerical examples and full wave simulation results are provided in section III. And section IV describes the 460 mm × 460 mm manufactured RA and the relevant experimental results at 26.5 - 28.5 GHz. Summary and conclusions are given in section V.

## II. DESIGN FLOW AND THEORIES

A RA is commonly used in the far field [20], while for the CATR application, it is actually more challenging in the near field. The near field of the RACATR system should satisfy the requirement for the quasi plane wave. In other words, the little rippled quiet zone can be attributed to desire for the ultra-low side lobe RA. As shown in Fig.1, firstly, the facility layout is required to be carefully designed to suppress the interference in the near field. Secondly, the design flow can optimize the aperture phase to control the near field diffraction and to synthesize the plane wave. Thirdly, the selection of the unit cell should be able to provide the phase shift. The manufactured RA

Manuscript received April 15, 2019; this work was supported by the National Natural Science Foundation of China under Grant No. 61671033.

All the authors are with the School of Electronics and Information Engineering, Beihang University, Beijing 100191, China (e-mail: lzp@buaa.edu.cn).

acts as a collimating device in the CATR system. Furthermore, the wideband formulas, which have been used to realize the wideband Hologram CATR in [9], is also applied in this system.

The geometry of a RACATR is illustrated in Fig. 2 [21]. The square aperture of RA is rotated 45 degrees around $z$ axis. The RA is illuminated by an offset feed horn at offset distance $\delta_x$ and the focal length $F$. The outgoing plane wave emanates at the off-axis angle $\theta$ at the same side of the feed horn to control the mirror-like reflection interference far away from the quiet zone. Then the quiet zone is generated at the propagation distance $h$ and behind the feed for reducing the feed leakage and avoiding the feed blockage.

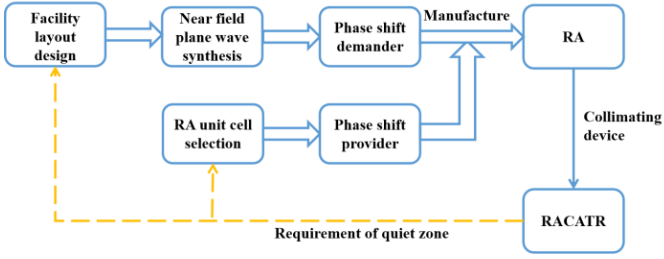

Figure 1. Design flow for the presented RACATR.

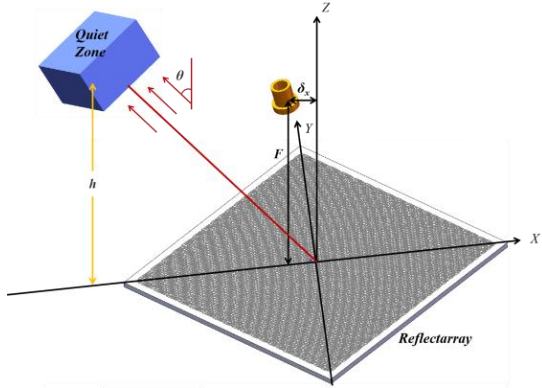

Figure 2. Facility layout for the presented RACATR.

*A. Facility layout design considerations*

The facility layout of the RACATR follows two golden key principles: reducing the edge diffraction and excluding the mirror-like reflection. The edge diffraction can be reduced by increasing the field gain in general. The mirror-like reflection appears when the feed illuminates the RA and a quasi-spherical wave front is re-radiated from the image point in the near field. This reflection is an inherent phenomenon caused by the ground of the RA and has to be avoided in the plane wave field. In the RACATR design, this effect has to be treated carefully because the desired quiet zone quality can be easily destroyed by the reflection interference, as shown in section III. Therefore, the offset distance $\delta_x$, the focal length-to-diameter ratio ($F/D$), the propagation distance $h$ and the plane wave outgoing angle $\theta$ should be properly designed to avoid the interference from the reflection area as illustrated in Fig. 3. The mirror-like reflection outgoing angle $\alpha$ has geometric relations with them as follows:

$$\tan(\alpha) = \frac{D/2 - \delta_x}{F} = \frac{1}{2 \times (F/D)} - \delta_x / F \quad (1)$$

$$D/2 + h \cdot \tan(\alpha) < h \cdot \tan(\theta) \quad (2)$$

$$h > F \quad (3)$$

where $F$ is the focal length, and $D$ is the diagonal length of the square.

The mirror-like reflection outgoing angle $\alpha$ decreases when the $F/D$ and offset distance $\delta_x$ increase as in (1). However, $F$ and $\delta_x$ also need to be controlled in the consideration of compactness and avoiding the feed blockage as well. The quiet zone locates out of the mirror-like reflection area as in (2). Then $h$ is larger than $F$ for reducing the feed leakage in (3). In other words, the facility layout parameters of the RACATR ($\delta_x$, $F/D$, $h$ and $\theta$) should be balanced to exclude the mirror-like reflection. As an example, a proper set of parameters ($\delta_x$ values 0.065 m, $F/D$ values 1.8, $h$ values 1.3 m and $\theta$ values 35°) are employed in the following simulations and the experiment.

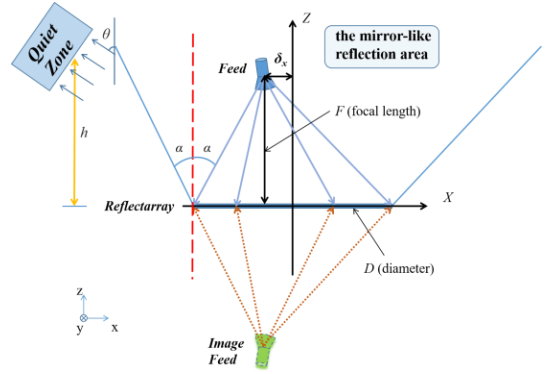

Figure 3. Top view of the RACATR and the designed layout apart from the mirror-like reflection.

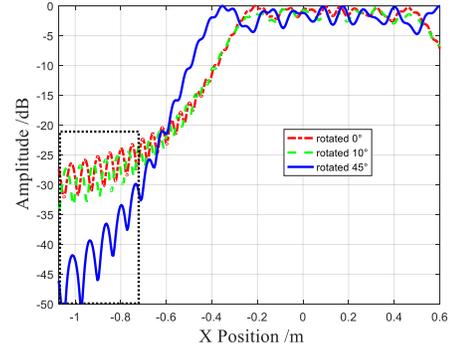

Figure 4. Scattering level of a metal square with the different rotated angles.

The off-axis distributed quiet zone, the longer focal length and the offset feed are aimed at suppressing the structural mirror-like reflection. A rotated square aperture also can reduce the unwanted diffraction from the truncated aperture and the physical installation frame. A metal square plate with the same layout as mentioned above is simulated to demonstrate that this facility layout of the RACATR can reduce the scattering energy in the near field of the square aperture, as shown in Fig. 4. The scattering level into the quiet zone is denoted in a dashed box. Obviously, the square aperture of RA, which is rotated 45°, helps the reduction of the scattering energy of the edge into the quiet zone. The structural scattering does not behaves as a sharp sinc function like the far field. The scattered beam is wide in the near field, and the levels are high on the cut lines parallel to the edge. So the 45° rotated angle of the aperture can reduce the scattering level to the minimum in the quiet zone. The field that located on the normal direction will be steered and synthesized into the plane wave field by the RA. The beam can be attributed to the resonant mode and terminal mode from the element antenna scattering [19]. To achieve an acceptable quality of the quiet zone, the desired beam has to be 20dB larger than the undesired ones. Both the edge diffraction and the mirror-like reflection are reduced through these optimized facility layout to enhance the quality of the quiet zone.

*B. Aperture design flow*

The basic principle of CATR is the use of a collimating device to generate the plane wave without requiring the normal far-field separation [1]. In the RACATR system, the RA acts as a plane wave generator in the Fresnel region of near field. The amplitude of the reflectarray aperture depends on the feed illumination, which is determined as the facility layout is fixed. As a feature of RA, the phase of each element is the only variable to be controlled for generating a quasi-plane wave field, which can be simplified as a plane wave synthesis problem of an only phased aperture for the near field application. The plane wave spectrum (PWS) theory as the base of the bidirectional propagation between the aperture and quiet zone is embedded into Intersection Approach (IA) to realize the aperture design [21-23].

The IA was developed to synthesize the antenna far field pattern at first [24], since Fourier transform pairs act as the shortcut between the source and the far field. As for the near field synthesis, Fourier transform pairs, which create the connection between the spatial domain and the angular spectrum domain in the near field, cannot be used directly. Fortunately, the PWS theory provides an analytical near field system transform function [21-23] to characterize the propagation in the angular spectrum domain. Employing the Fourier transform pairs and PWS theory, a round trip from aperture field to near field is established, which can be embedded into IA for the plane wave synthesis in the near field.

The aperture design flow for RACATR is illustrated in Fig. 5. Firstly, the initial amplitude and phase of the aperture are calculated by the feed radiation pattern based on the proposed facility layout with the offset feed and the off-axis outgoing wave. The amplitude is constrained on a constant illumination while the phase can be adjusted freely in the optimization. The complex aperture field is transformed to the aperture angular spectrum by an IFFT. Multiplying the near field system transform function provided by the PWS theory in the angular spectrum domain, the field propagates from the aperture to the near field. Then the near field with amplitude and phase in the spatial domain is obtained by an FFT of the near field angular spectrum. This is the connection bridge between the aperture and the near field. An area of the near field is selected as the quiet zone whose amplitude and phase are optimized, shaped and desired. Through the inverse steps, the modified near field can be transformed back to the aperture field, whose phase will be utilized into the next iteration with the constant constrained amplitude Through multiple iterations in a few minutes, the synthesized near field meets the requirement of the quiet zone field distribution (typically less than ±1dB in amplitude ripple and ±10° in phase ripple).

Of course, other traditional global optimization methods also can be used for the design and may cost more time, but this method is recommended for the light computation and fast speed, which can be completed for the several thousands of optimized variables in a few minutes.

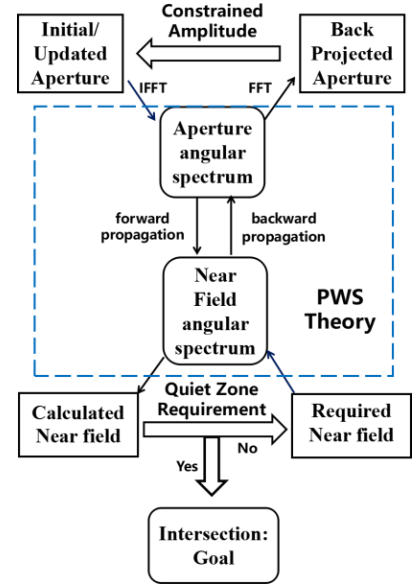

Figure 5. The only phased aperture design flow of the RACATR.

*C. The unit cell*

To obtain a RA that is able to generate a plane wave in certain frequency bandwidth, the phase shift of the unit cell should cover more than 360° and the element is supposed to own the wideband performance. In order to reduce the impact of fabrication errors, maximum phase sensitivity with respect to the element size change is better to be less than 200°/mm [25].

The Phoenix cell is selected as the phased element of this RA. With the unique rebirth capability, the Phoenix cell is able to achieve a full 360° phase shift cover [8]. The geometry of the Phoenix cell for the RA is shown in Fig. 6. The middle ring grows between the fixed outer ring and center patch. When the middle ring is the same size of the outer ring, the element returns to the initial state just like the rebirth of the phoenix. The element is designed for printing on Rogers 4350 substrate with a dielectric constant of 3.48. The design of a thick air layer [25] reduces the effective dielectric constant and improves the wideband performance of the element. By adjusting the parameters of element, the maximum phase sensitivity is less than 150°/mm. The values of other geometrical parameters is given in Table I. The element reflection phase versus $L_r$ (the side length of the middle square ring) at 26.5-29.5 GHz is shown from top to bottom in Fig. 7 simulated by FEKO Suite 7.0. The element reflection phase curves show the linear characteristic, which ensures a stable reflection phase in the wide frequency band and reduces the impact of manufacture errors.

TABLE I
GEOMETRICAL PARAMETERS OF THE PHOENIX CELL

| Parameters | $L_{in}$ | $L_{out}$ | $L_r$ | $w$ | $h_1$ | $h_2$ |
|---|---|---|---|---|---|---|
| Values [mm] | 1 | 5 | 1~5 | 0.15 | 1.8 | 0.762 |

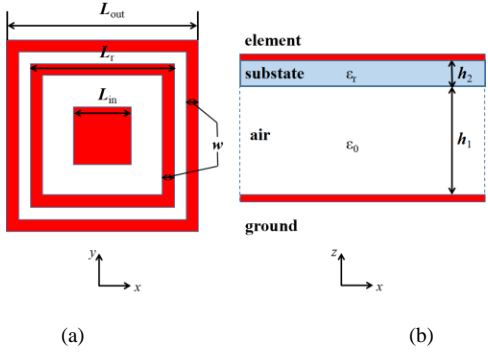

(a) (b)
Figure 6. The Phoenix cell. (a) Top view. (b) Side view.

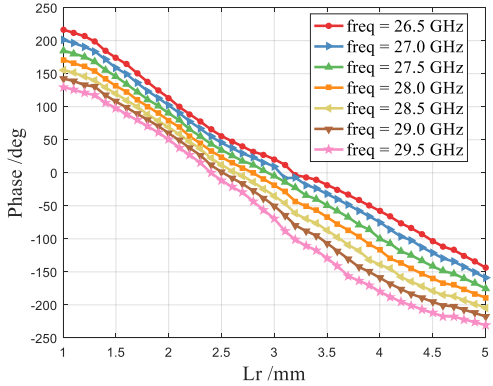

Figure 7. The reflection phase versus $L_r$ at 26.5-29.5 GHz.

*D. Wideband formulas*

For a printed micro-strip RA, the bandwidth is primarily limited by two factors [4]: the element and the differential spatial phase delay. The wideband Phoenix cell with thick air layer relieves the first factor. And the spatial phase delay can be compensated by linearly adjusting the feed location [9], which is demonstrated as follows.

The center of the unit cell locates any point $(x,y)$ on the reflectarray aperture, and the wave path from the feed $c(x,y)$ should be collimated into the desired plane wave path:

$$c(x, y) = \sqrt{(x+\delta_x)^2 + y^2 + F^2} + (x+\delta_x)\sin\theta \quad (4)$$

$$c(x, y) = F \cdot \sqrt{\frac{(x+\delta_x)^2 + y^2}{F^2} + 1} + (x+\delta_x)\sin\theta \quad (5)$$

Where $\theta$ is the plane wave off-axis angle, and $\delta_x$ is offset position of the feed for outgoing apart from the structural mirror-like reflections sourced from the RA unit cells and the installation frames.

If focal length $F$ is much bigger than $\sqrt{(x+\delta_x)^2 + y^2}$, Taylor series expansion can be used to simplify the formula (5).

$$c(x, y) \approx F \cdot \left[1 + \frac{1}{2} \cdot \frac{(x+\delta_x)^2 + y^2}{F^2}\right] + (x+\delta_x)\sin\theta \quad (6)$$

The phase of the points $(x,y)$ are $\varphi(x,y)$ at frequency $f$.

$$\varphi(x, y) = \frac{2\pi f}{c} \cdot \left[F + \frac{1}{2} \cdot \frac{(x+\delta_x)^2 + y^2}{F} + (x+\delta_x)\sin\theta\right] \quad (7)$$

The phase of the points $(x,y)$ are $\varphi'(x,y)$ at a new frequency $f'$. And $f'=M \cdot f$. The zero order component is constant. Linearly adjusted the feed location to the new focal length $F'=M \cdot F$. The first order component of the phase $\varphi'(x,y)$ are the same as $\varphi(x,y)$. Moreover, if the plane wave outgoing angle $\theta$ become new angle $\theta'$ and $\sin(\theta')=M \cdot \sin(\theta)$, the differential spatial phase delay will disappear at different frequencies. However, this method is disabled when the focal length $F$ is not much bigger than $\sqrt{(x+\delta_x)^2 + y^2}$ and the high order component cannot be ignored for the higher frequencies.

In Fig. 8, a RA is offset illuminated by the feed at offset distance $\delta_x$ for generating the plane wave at the center frequency $f$. And the initial feed location is $F$. The RA is located in the $xy$-plane and the plane wave emanates at angle $\theta$. For the wideband application, the relationship at another frequency $f'$ can be formulated as demonstrated above:

$$\begin{cases} F' = \dfrac{f'}{f} F \\ \sin(\theta') = \dfrac{f}{f'}\sin(\theta) \end{cases} \quad (8)$$

For a higher frequency $f'$, the feed should be moved further away to the feed location $F'$ and the plane wave is emanating at angle $\theta'$ to the new quiet zone. However, if the feed is moved too far away at higher frequencies, the edge is illuminated at the higher level while the mirror-like reflection becomes far away from the quiet zone. For a lower frequency, the feed moves closer to the RA. The mirror-like reflection angle $\alpha$, shown in Fig. 3, increases as feed location moves forward to the RA. The interference of the reflect wave raises significantly as the feed is too close to the RA.

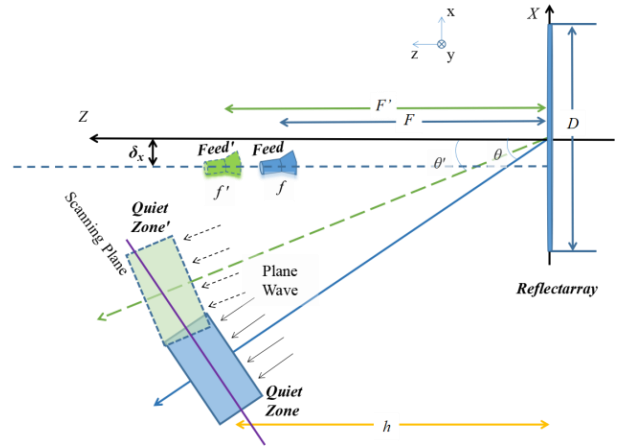

Figure 8. Top view of the RACATR for the wideband application.

## III. NUMERICAL SIMULATION

In this part, the ideal aperture is optimized by the presented design approach to verify the feasibility of the RACATR. The performance of the linear reflectarray is compared with the traditional reflector layout to ensure that this facility layout can prevent the quiet zone from the reflection interference, in which the structural reflections from the physical structure of RA has to be included in the model. Moreover, the tolerances are also assessed on the basis of the full wave simulation results.

### A. Ideal aperture

In consideration of the manufacturing capability of PCB and the practicability of the RACATR, the side length of the RA is designed as 460 mm at 28 GHz. However, the two dimensional RA is too large to be simulated by the full wave method in the commercial software.

As an idealized and simplified model of the RA, the aperture is amplitude-constrained and the only phase is optimized for generating the plane wave, which is illustrated in Fig. 9. The spacing within cells is 5 mm. And the only-phased aperture is defined as a 92 × 92 ideal point source array. The amplitude is determined by the feed pattern, the feed location $F$ and the offset distance $\delta_x$, shown in Fig. 10. The propagation height is $h$. And the plane wave emanates at the angle $\theta$ and the quiet zone plane is parallel to the aperture plane with a size of 0.23 m × 0.23 m. The value of geometrical parameters are given in Table II.

The design approach is presented in the aperture design flow as mentioned. Through 1000 times iterations in a few minutes, the initial phase that only compensates the phase difference for the off axis angle $\theta$ is converged to the optimized phase shown in Fig. 11. The performances of quiet zones are compared in Fig. 12. The ripple (half of the peak-to-peak value) of the amplitude was reduced from 3.12 dB to just over 0.14 dB and the ripple of phase was from 7.03° to 1.06° after removing the slope, which fulfills the traditional requirement of the quiet zone at 28 GHz.

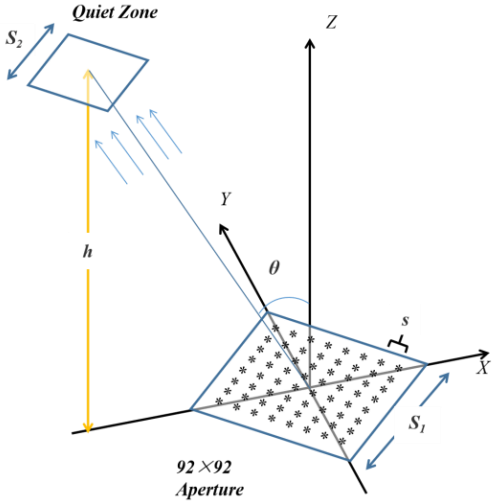

Figure 9. The only phased aperture with the geometrical parameters.

TABLE II
GEOMETRICAL PARAMETERS OF THE ONLY PHASED APERTURE

| Parameters | $s$ | $h$ | $S_1$ | $S_2$ | $F$ | $\delta_x$ | $\theta$ |
|---|---|---|---|---|---|---|---|
| Values [m] | 0.005 | 1.3 | 0.46 | 0.23 | 1.21 | 0.065 | 35° |

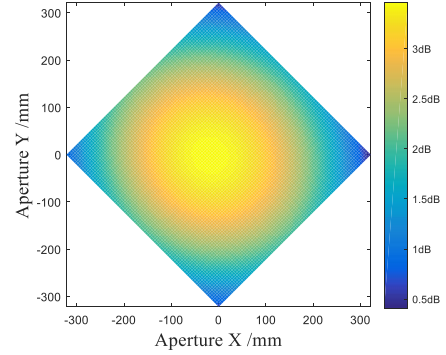

Figure 10. The constrained amplitude by the feed illumination in the only phased aperture (the amplitude peak is offset in *x*-direction).

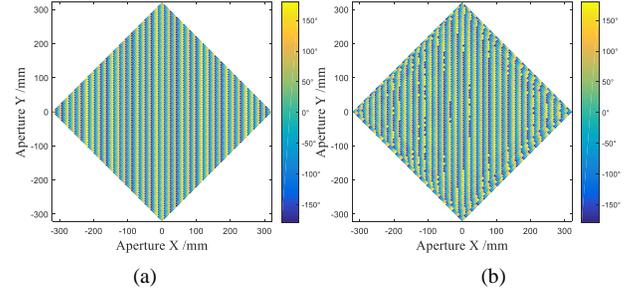

(a)          (b)

Figure 11. The phase of the aperture. (a) Initial. (b) Optimized.

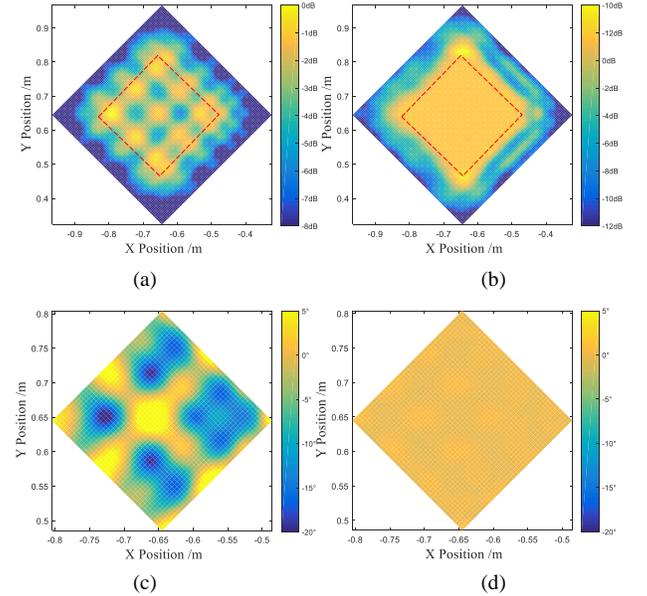

Figure 12. The performance of the quiet zone. (a) Initial amplitude. (b) Optimized amplitude. (c) Initial phase. (d) Optimized phase.

The wideband formulas are also verified in the numerical examples. The calculation flow of virtual aperture phase at different frequencies or feed locations is presented in Fig. 13. In order to compensate the phase delay caused by the feed illumination at 28 GHz, the aperture phase $\Phi_a(x,y)$ subtracts the illuminated phase $\Phi_b(x,y)$ and the result is the element phase $\Phi_e(x,y)$, the reflection phase that reflectarray elements are supposed to provide. Then the reflection phase versus $L_r$ curve at 28 GHz is utilized to obtain the reflectarray model $L(x,y)$. The $L(x,y)$ is the constant model element set on the RA. For these fixed elements $\Phi'_e(x,y)$ are recalculated at other frequencies. Through the inverse process, the new aperture phase $\Phi'_a(x,y)$ can be calculated. Note that the illuminated phase $\Phi'_b(x,y)$ is also changed with the working frequency and the feed location. Once the phase and the amplitude of aperture are decided, the quiet zone field can be recomputed.

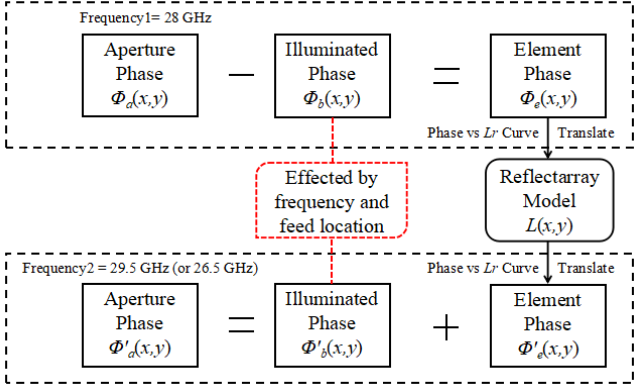

Figure 13. The calculation flow of aperture phase.

The wideband application results are summarized in Table III. The plane wave outgoing angle is 34.09° and has little difference with the target value 35°, which is the consequence of the plane wave synthesis. When the working frequency is changed without moving the feed, the ripple of amplitude and phase deteriorates as the spatial phase is mismatched. However, when the feed location adjusts as the wideband formulas predicted for the different frequencies, the plane wave outgoing angle $\theta$ changes as well. And the quiet zone performance improves a lot at 26.5 GHz. The situation at higher frequency has not improved significantly because the high order component at the higher frequencies becomes larger and cannot be compensated as proposed above.

TABLE III
WIDEBAND APPLICATION OF THE ONLY PHASED APERTURE

| Frequency [GHz] | Feed Location $F$ [m] | Quiet Zone Ripple | | $\theta$ [deg] |
|---|---|---|---|---|
| | | Amplitude [±dB] | Phase [±deg] | |
| 26.5 | 1.142 | 0.279 | 1.30 | 35.97 |
| 26.5 | 1.207 | 0.700 | 6.40 | 36.73 |
| 28.0 | 1.207 | 0.144 | 1.06 | 34.09 |
| 29.5 | 1.207 | 0.798 | 8.35 | 31.85 |
| 29.5 | 1.271 | 0.654 | 3.79 | 32.15 |

*B. Linear reflectarray simulation*

The numerical examples, based on the calculation of the ideal aperture, verify the design flow of the RACATR. However, the actual reflectarray unit cells have the solid structure and the complex reflection or interaction, which is simplified and ignored in the numerical examples (section A). In consideration of the feasibility of full wave simulation, the linear reflectarray is simplified to imitate the planar one which is too computationally heavy to be simulated. In this way, the performance of the linear reflectarray is compared with the ideal aperture to include the solid structural reflection and to check the designed layout.

The linear reflectarray with 130 cells is illustrated in Fig. 14. The length of the linear reflectarray is 0.65 m, the same length as the diagonal line of the planar reflectarray. Other geometrical parameters of the RACATR facility layout, such as the feed location $F$, the offset distance $\delta_x$, the propagation height $h$ and the off axis angle $\theta_1=35°$, are the same as provided above. The quiet zone and the feed locate at the same side of the normal of the aperture in the facility so as to depart from the structural reflections. The second linear reflectarray has the offset angle $\theta_2=20°$ and the outgoing angle $\theta_2=20°$, as employed in [14-18] to imitate the traditional reflector CATR facility layout. Besides, two ideal linear apertures with $130 \times 1$ point source array are independently optimized for the better quiet zone. The optimized phase settings in different facility layouts are employed both for the linear reflectarray models.

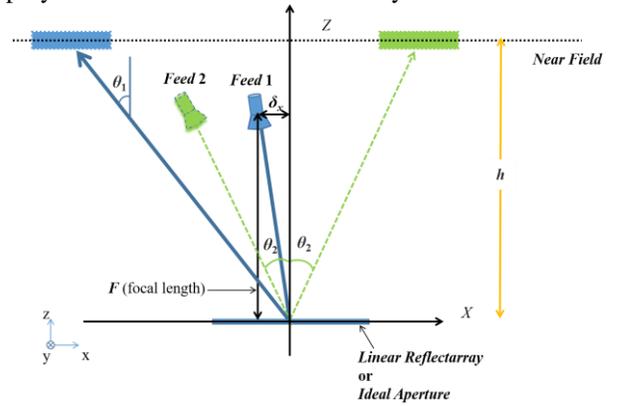

Figure 14. The linear reflectarray with two geometrical layouts.

The simulated amplitude of quiet zone are shown in Fig. 15. The little ripple are generated by the ideal apertures in different facility layouts between the dashed edge lines. The amplitude ripple of the first linear reflectarray model is under ±1 dB and the performance is similar to the ideal aperture, whose facility layouts are the RACATR type as proposed above. The shape of the amplitude performance is the consequence of the aperture design flow in the condition of this facility layout. The power level of the linear reflectarray model is lower than the ideal aperture in the quiet zone, because the reflection interference in the mirror-like reflection area takes away part of the energy. But the power of the reflection interference in the quiet zone is too low to disturb the performance thanks to that this facility layout puts the plane wave field out of the mirror-like reflection. The still existing difference between the ideal aperture and the linear reflectarray is sourced from the diffraction effect of the mirror and from the mutual coupling of reflectarray elements. However, the performance of the second linear reflectarray cannot be accepted because of the great difference from the ideal aperture even with the same facility layout. The directly imitating the facility layout of the reflector CATR in the

RACATR makes the structural mirror reflection seriously into the quiet zone. The peak level of the structural reflection is only 10dB lower than the desired plane wave, shown in Fig.15 and 16. The mirror reflection behaves as the diffused spherical wave front interfered with the quiet zone field. The high energy of the mirror reflection has to be abandoned even at the cost of the lower efficiency.

The co-polarization and the cross-polarization results of the first linear reflectarray are provided in Fig. 16 and the difference is over 20 dB in the quiet zone field.

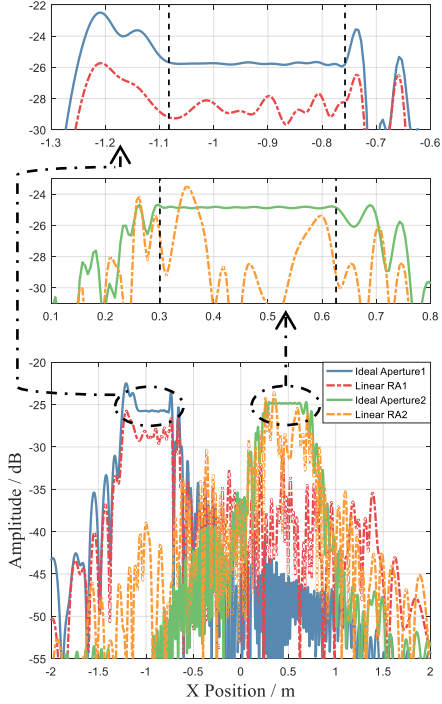
Figure 15. The amplitude of the linear reflectarrays and ideal apertures.

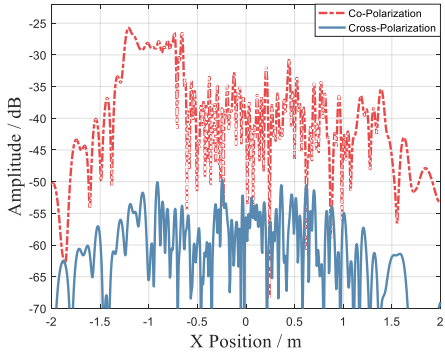
Figure 16. The amplitude of the co-polarization and the cross polarization

### C. The assessment on the tolerances

Based on the linear reflectarray full wave simulation, this part mainly analyzes the impact of some common error sources, including the manufacture errors of the cells and position errors of the feed. The manufacture errors of the cells mainly refer to the manufacture errors of the length of the middle ring $L_r$. Some random errors are set to all cells of the linear reflectarray and the results are shown in Fig. 17 and Table IV. The plane wave field locates between the dashed lines in the amplitude and the phase of the plane wave field is fitted by removing the slope.

The cells with $\pm \lambda/200$ or $\pm \lambda/100$ errors have a slight effect on the plane wave field performance. Obviously, when the cells have errors bigger than $\pm \lambda/50$, the ripple of amplitude is more than $\pm 1$ dB and the ripple of phase is more than $\pm 10°$. Because the manufacture errors bring the wrong phase drift randomly and the quality of the plane wave field turns bad as expected. Furthermore, the linear element reflection phase curves of the Phoenix cell also help to relax the deviation of the phase shift caused in the manufacture.

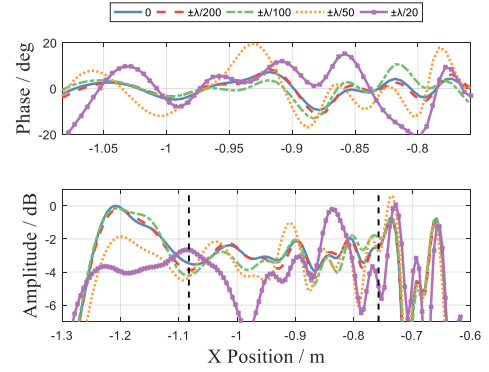
Figure 17. The quiet zone performance of the linear reflectarray with the different errors of the cells.

TABLE IV
THE QUIET ZONE PERFORMANCE INFLUENCED BY MANUFACTURE ERRORS

| Manufacture errors | 0 | $\pm \lambda/200$ | $\pm \lambda/100$ | $\pm \lambda/50$ | $\pm \lambda/20$ |
|---|---|---|---|---|---|
| Amplitude ripple [±dB] | 0.99 | **1.19** | 1.52 | 2.53 | 3.65 |
| Phase ripple [±deg] | 8.28 | **9.83** | 11.81 | 18.27 | 17.78 |

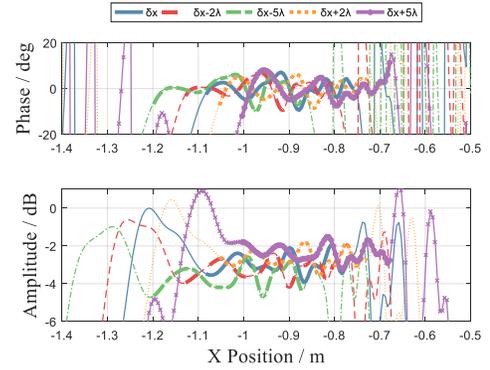
Figure 18. The quiet zone performance of the linear reflectarray with different horizontal position errors of the feed.

TABLE V
THE QUIET ZONE PERFORMANCE INFLUENCED BY TRANSVERSE OFFSET OF FEED

| Transverse position | $\delta_x+5\lambda$ | $\delta_x+2\lambda$ | $\delta_x$ | $\delta_x-2\lambda$ | $\delta_x-5\lambda$ |
|---|---|---|---|---|---|
| Predicted $\theta$ [deg] | 32.55 | 34.02 | 35.00 | 35.98 | 37.45 |
| Actual $\theta$ [deg] | 32.22 | 33.74 | 34.79 | 35.86 | 37.49 |
| Amplitude ripple [±dB] | 0.86 | 0.91 | 0.99 | 0.90 | 1.06 |
| Phase ripple [±deg] | 11.75 | 7.39 | 8.28 | 8.05 | 8.01 |

The position errors of the feed divides into two directions for the linear reflectarray, the offset in the transverse and the down direction. The transverse offset distances $\delta_x$ for ±2λ and ±5λ are simulated and the results are compared in Fig. 18 and Table V. According to the formula (7), when the offset distance $\delta_x$ increases, the value of the third component will follow to increase and then the plane wave outgoing angle $\theta$ can drop to keep the balance. The difference value of the plane wave outgoing angle $\theta$ can be simply predicted by the arc tangent value of the ($\Delta/F$), where $\Delta$ is the small transverse deviation. And the actual value of the plane wave outgoing angle $\theta$ is calculated by the slope of the phase result. Those values are summarized in the Table IV and the actual values agree well with the predicted values. When the plane wave outgoing angle $\theta$ decreases, the plane wave field locates closer to the axis and the electrical level will rise as the projection from the linear reflectarray is stronger. Both the amplitudes and the phases of different conditions in their plane wave fields are marked in the bold line and the amplitudes perform as expected above. The performance of the linear reflectarray with ±2λ transverse errors fulfills the requirement of the quiet zone. When the errors increase to ±5λ, the quality of the quiet zone goes bad. However, the reflection interference is stronger as the plane wave field locates too close to the axis and the desired plane wave field can be easily destroyed.

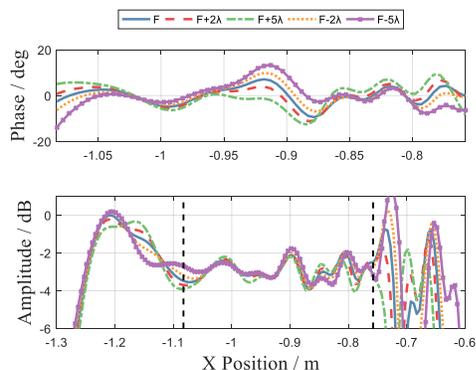

Figure 19. The quiet zone performance of the linear reflectarray with different down position errors of the feed.

TABLE VI
THE QUIET ZONE PERFORMANCE INFLUENCED BY DOWN OFFSET OF FEED

| Down position | F+5λ | F+2λ | F | F-2λ | F-5λ |
|---|---|---|---|---|---|
| Amplitude ripple [±dB] | 0.95 | 0.95 | 0.99 | 0.99 | 0.961 |
| Phase ripple [±deg] | 10.88 | 8.99 | 8.28 | 8.47 | 13.60 |

The down offsets of the feed location are set for ±2λ and ±5λ, and the simulated results are presented in Fig. 19 and Table VI. The performance of the linear reflectarray with ±2λ down errors fulfills the requirement of the quiet zone. When the errors increase to ±5λ, the quality of the quiet zone goes bad. According to the formula (7), the plane wave outgoing angle $\theta$ has little relations with the down errors, so the quiet zone locations are almost the same as in Fig. 19. When the down feeding location increases, the phase will increase but the speed of increasing is slower at the place where is closer to the feed center. The phase is a relative value, so the phase in the center relatively decreases and forms a valley. In contrast, the peak of phase will be formed when the down offset decreases. The amplitudes are similar to each other and the phases perform as expected above.

## IV. EXPERIMENTS

An experimental RACATR is developed with the dimensions of 460 mm × 460 mm. The experiment system is shown in Fig. 20, which consists of a RA, a feed, a test cable, a network analyzer, a planar scanner, and some absorbing materials. Two Ka-band six-slots horns are employed as the feed and the scanning probe respectively. The layout parameters of the experiment system such as the reflectarray manufacturing parameters $L(x,y)$, the feed location $F$, the propagation distance $h$, the offset distance $\delta_x$ and the plane wave outgoing angle $\theta$ are the same as those of the numerical examples in section III. The manufactured array by PCB is installed on a metal frame.

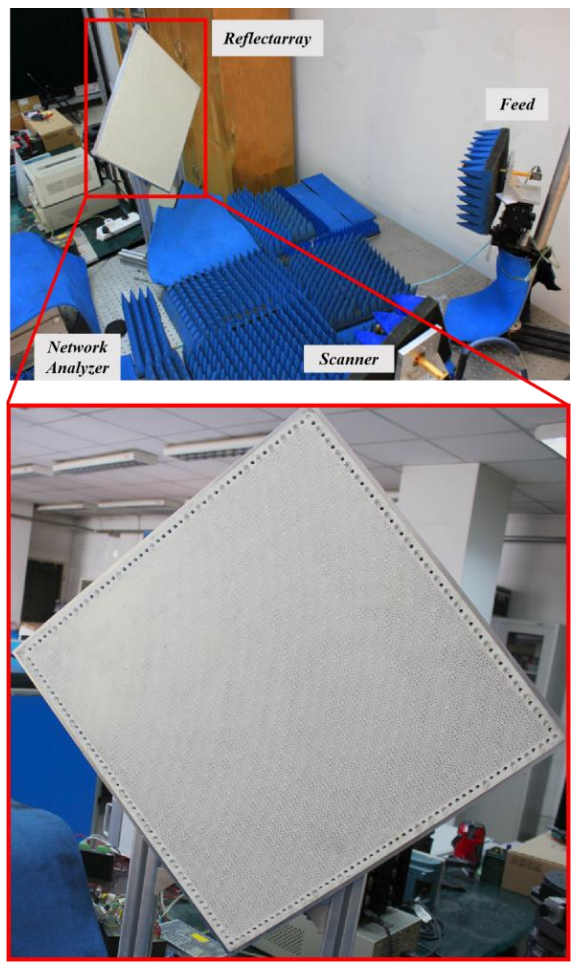

Figure 20. The experiment layout of the RA CATR.

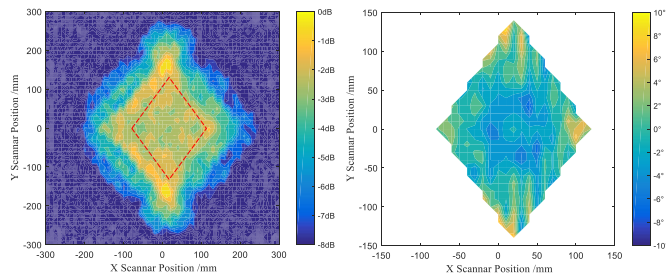

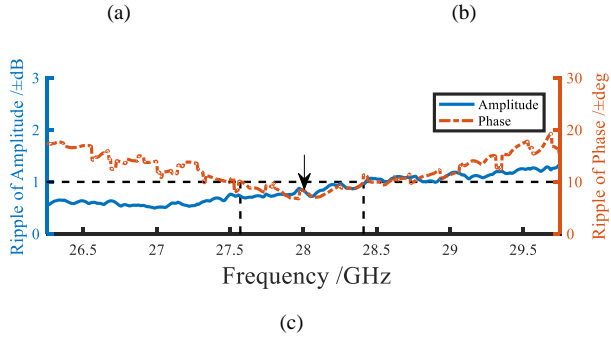

Figure 21. The performance of the scanning plane at the feed location $F_1$. (a) the amplitude and (b) the phase of the quiet zone for 28 GHz, (c) the wide band performance of the RACATR (Feeding location 1.207m).

The performance of the scanning plane at the feed location $F_1$=1.207 m is shown in Fig. 21. The quiet zone size in the X axis is shorter than the one in the numerical examples, because the scanning plane is vertical to the propagation direction of the outgoing plane wave rather than parallel to the reflectarray aperture. The ridge-like area of the quiet zone is formed in the scanning plane for two reasons. Firstly, it is the consequence of the plane wave synthesis. The peaks, which locate at the edge of the quiet zone, are the same as shown in Fig. 12 (b). Secondly, the ridges are relevant to the reflection from the metal structure (include in the frame) on the reflectarray aperture especially in the left side, because their illumination level from the feed is higher. The quiet zone is denoted in the dashed box with 200 mm in horizontal and 290 mm in vertical direction. Considering the plane wave outgoing angle $\theta$, the use ratio of the aperture is 53% in horizontal size and 43% in vertical direction. The phase performance of the quiet zone is shown in Fig. 21 (b) by removing the slope. The ripple (half of the peak-to-peak value) of the amplitude is 0.83 dB and the ripple of phase is 8.35°, which fulfills the requirement of the quiet zone at 28 GHz and agrees well with the linear reflectarray simulation results. The bandwidth performance of the RACATR is also provided in Fig. 21 (c). In consideration of the quiet zone requirement, the RACATR can be directly applied from 27.57 GHz to 28.41 GHz with 3% relative bandwidth. This bandwidth maybe enough for one practical test, which requires RF bandwidth for 800 MHz in [26]. Furthermore, the plane wave outgoing angle is 33.93° and has little difference with the target value 35°, which is also the consequence of the plane wave synthesis and agrees well with the ideal aperture performance.

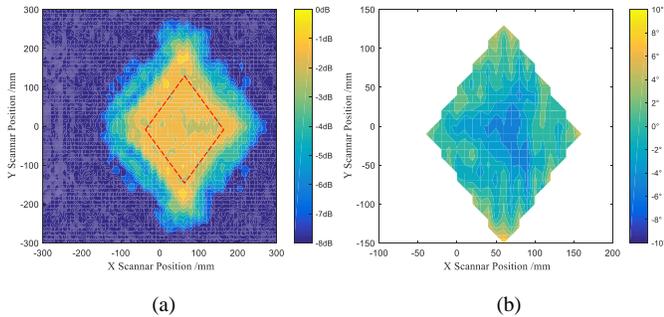

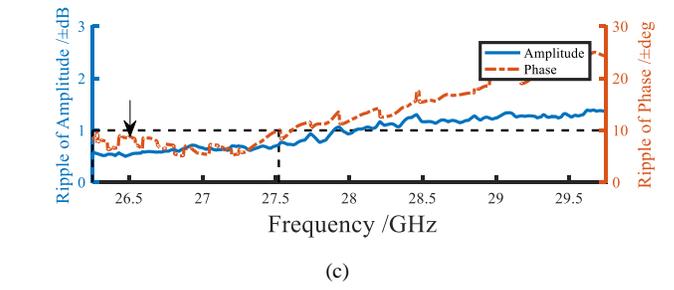

Figure 22. The performance of the scanning plane at the feed location $F_2$. (a) the amplitude and. (b) the phase of the quiet zone for 26.5 GHz, (c) the wide band performance of the RACATR (Feeding location 1.142m).

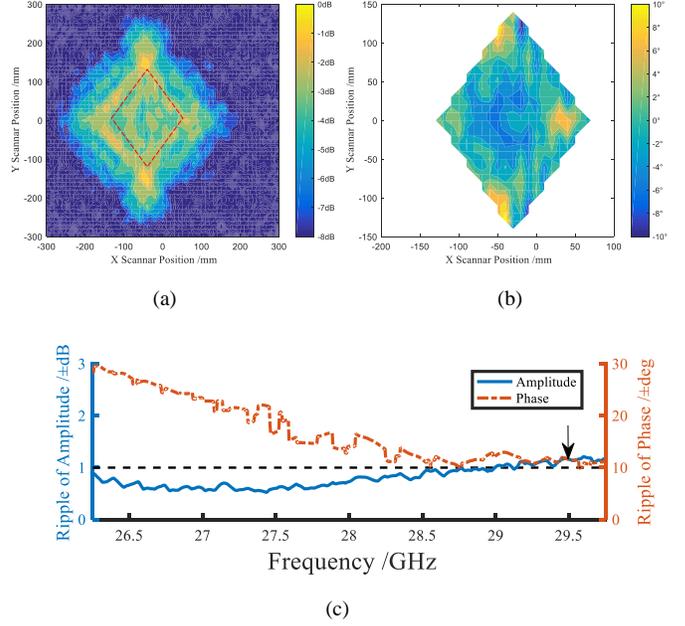

Figure 23. The performance of the scanning plane at the feed location $F_3$. (a) the amplitude and (b) the phase of the quiet zone for 29.5 GHz, (c) the wide band performance of the RACATR (Feeding location 1.271m ).

The RACATR can expand the working frequency by linearly adjusting the feed and the wideband formulas are verified with three feed locations experimentally. Three feed locations are calculated by the wideband formulas based on the feed location $F_1$ at the center frequency, where $F_1$ is 1.207 m for 28 GHz, $F_2$ is 1.142 m for 26.5 GHz and $F_3$ is 1.271 m for 29.5 GHz. For the feed location $F_2$=1.142 m, the ripple of the amplitude is 0.53 dB and the ripple of phase is 8.86° at the matched frequency 26.5 GHz in Fig. 22. Linearly adjusting the feed location from $F_1$ to $F_2$ makes the quality of the quiet zone successfully meet the requirement. In Fig. 22 (c), the bandwidth performance of the RACATR is also provided. The RACATR can be directly applied from 26.25 GHz to 27.52 GHz. The wideband formulas are effective for the lower frequency. However, for the feed location $F_3$=1.271 m, the ripple of the amplitude is 1.16 dB and the ripple of phase is 11.50° at the matched frequency 29.5 GHz in Fig. 23. The result does not meet the requirement but has a tendency to improve at 29.5GHz.

The measured results of the wideband application for RACATR are summarized in Table VII. The plane wave outgoing angle $\theta$ agrees well with the ideal aperture. And the quality of the quiet zone is close to the linear reflectarray simulation results. When the feed location is changed from $F_1$ to $F_2$, the quiet zone performance at 26.5 GHz can be improved. For the situation at 29.5 GHz, the improvement is not obvious mainly for two factors: The first is that the phase compensation by linearly adjusting the feed location plays a limited role at higher frequencies because the high order component at the higher frequencies becomes larger and cannot be compensated as mentioned in numerical example. The other is that the plane wave outgoing angle $\theta$ is smaller at higher frequencies, therefore, the reflection interference is nearer to the quiet zone. In other words, the design frequency of a RACATR aperture is better to choose a higher value than the average value of the band and use the lower frequency in the test as far as possible.

TABLE VII
WIDEBAND APPLICATION OF THE RACATR

| Frequency [GHz] | Feed Location $F$ [m] | Quiet Zone Ripple | | $\theta$ [deg] |
|---|---|---|---|---|
| | | Amplitude [±dB] | Phase [±deg] | |
| **26.5** | $F_2$=1.142 | **0.53** | **8.86** | **36.12** |
| 26.5 | $F_1$=1.207 | 0.63 | 16.98 | 36.74 |
| **28.0** | $F_1$=1.207 | **0.83** | **8.35** | **33.93** |
| 29.5 | $F_1$=1.207 | 1.25 | 17.47 | 31.48 |
| **29.5** | $F_3$=1.271 | **1.16** | **11.50** | **31.46** |

The cross-polarization of the RACATR is also tested for the whole quiet zone at the feed location $F_1$=1.207 m. The global maximum value of the cross-polarization in the quiet zone is −21.36 dB at 28GHz.

## V. CONCLUSION

In this work, a RA type of CATR is proposed. The new facility layout, including the offset feed, the same-side off-axis quiet zone and 45 degree rotated aperture, is utilized to reduce the disturbing components for generating the quiet zone. This novel method employs an IA-synthesized RA to generate the desired plane wave field. The phoenix cell and the adjusting feed location are recommended for the wide band realization of the RA CATR. As an easier to be manufactured and lower-cost alternative for the conventional reflector compact range, the RA is demonstrated at 26.5~28.5GHz and can be utilized for 5G millimeter wave antenna measurement. Moreover, this RA type of CATR has a broad prospect for the application in automotive radar antenna measurement at 77 GHz or even at terahertz band.